\documentclass[a4paper,prl,superscriptaddress,twocolumn]{revtex4}

\usepackage{amsmath,bm}

\newcommand{\ccfont}[1]{\textbf{#1}}
\newcommand{\poly}{\mathrm{poly}}
\newcommand{\tr}{\mathrm{tr}}
\newcommand{\QMA}{\ccfont{QMA}}
\newcommand{\NP}{\ccfont{NP}}
\newcommand{\MA}{\ccfont{MA}}
\renewcommand{\P}{\ccfont{P}}
\newcommand{\coNP}{\ccfont{coNP}}
\newcommand{\ket}[1]{|#1\rangle}
\newcommand{\bra}[1]{\langle#1|}

\begin{document}

\title{The computational difficulty of finding MPS ground states}

\author{Norbert Schuch}
\affiliation{Max-Planck-Institut f\"ur Quantenoptik, 
	Hans-Kopfermann-Str.\ 1, D-85748 Garching, Germany.}
\author{Ignacio Cirac}
\affiliation{Max-Planck-Institut f\"ur Quantenoptik, 
	Hans-Kopfermann-Str.\ 1, D-85748 Garching, Germany.}
\author{Frank Verstraete}
\affiliation{Fakult\"at f\"ur Physik, Universit\"at Wien,
Boltzmanngasse 5, A-1090 Wien, Austria.}

\begin{abstract}
We determine the computational difficulty of finding ground states of
one-dimensional (1D) Hamiltonians which are known to be Matrix Product
States (MPS).  To this end, we construct a class of 1D frustration free
Hamiltonians with unique MPS ground states and a polynomial gap above, for
which finding the ground state is at least as hard as factoring. By
lifting the requirement of a unique ground state, we obtain a class for
which finding the ground state solves an \NP-complete problem. Therefore,
for these Hamiltonians it is not even possible to certify that the ground
state has been found. Our results thus imply that in order to prove
convergence of variational methods over MPS, as the Density Matrix
Renormalization Group, one has to put more requirements than just MPS
ground states and a polynomial spectral gap.  
\end{abstract}

\maketitle

Explaining the behaviour of correlated quantum many-body systems is one of
the major challenges in physics and fundamental to the understanding of
condensed matter systems. The exponential dimension of the underlying
Hilbert space renders a straightforward numerical simulation impossible
both with respect to computational time and storage space.  However, for
specific physical scenarios simulation methods have been developed; in
particular, the Density Matrix Renormalization Group (DMRG)
method~\cite{white,schollwoeck} has proven extremely successful in
describing ground and thermal states of one-dimensional spin systems, and
to some extent even their time evolution, to very high accuracy.

DMRG has a natural interpretation as a variational ansatz over the class
of Matrix Product States (MPS)~\cite{frank:dmrg-mps}, and it has been
proven that every ground state of a local gapped Hamiltonian can indeed be
approximated efficiently by MPS~\cite{hastings:arealaw}.  While this
explains why MPS are well suited to describe ground states of
one-dimensional quantum systems, essentially nothing fundamental
concerning the convergence of the variational method over the class of MPS
could be shown, despite the fact that in practice, it converges extremely
well.  Actually, given the type of optimization problem at hand, one is
rather tempted to believe that DMRG will typically get stuck in local
minima, and in fact it has been shown that if the optimization is
performed in a specific way where sites are optimized simultaneously,
configurations might occur where the optimization problem becomes \NP\
hard, meaning the algorithm would get stuck~\cite{jens:dmrg-np}.  Yet, 
this difficulty is solely due to the specific way in which the
optimization is performed, rather than being a fundamental problem of any
variational method over MPS.

As we show in this paper, however, under natural assumptions on the
Hamiltonian obtaining a sufficiently good MPS approximation to the ground
state, and in particular finding the minimum in DMRG, is a computationally
hard problem, i.e., can in the worst case take exponential time.  More
precisely, we construct a class of nearest neighbor Hamiltonians on a
one-dimensional chain of length $L$ with the following properties:
Any of those Hamiltonians has a unique ground state with a spectral gap of
order $1/\poly(L)$ above, it is frustration free (i.e.\ the ground state 
minimizes each local term of the Hamiltonian), and the
ground state is a Matrix Product State of size polynomial in $L$, as are
the low-lying excited states.  For these Hamiltonians, we show that
finding the ground state (or a polynomial-accuracy approximation thereof)
is a hard problem, as it e.g.\ encompasses factoring numbers.  This
implies that finding an MPS approximation of these ground states most
likely cannot be solved efficiently by classical computers.  

By considering a slightly less restricted  class of Hamiltonians, we
obtain further results. In particular, if instead of requiring a unique
MPS ground state we allow for a ground state subspace spanned by MPS while
keeping the polynomial energy gap, and instead of frustration freeness
require the ground states only to be eigenstates to each local term, we
obtain a class of Hamiltonians for which finding the ground state is an
\NP-complete problem. This implies that for this class,
it is even impossible to certify that the ground state has
been found, based on widely believed complexity theoretic assumptions.
Moreover, it follows that there cannot even be a DRMG-like
algorithm which works at least for frustration-free systems, since this 
would allow for the efficient solution of \NP-complete problems.

Let us start by briefly introducing Matrix Product States (MPS) and their
entanglement properties. An MPS on a length $L$ chain of $d$-level
systems (``spins'') with \emph{bond dimension} $D$ is given by
\[
\ket\psi=\sum_{i_1,\dots,i_L}
    \tr[A^{[1]}_{i_1}\dots A^{[L]}_{i_L}]
    \ket{i_1,\dots,i_L}\ ,
\]
where the $A^{[k]}_{i_k}$ are $D\times D$ matrices. MPS satisfy the
entropic area law, i.e., the entanglement across each cut is bounded by
$\log D$, and they efficiently approximate ground states of gapped local
Hamiltonians~\cite{hastings:arealaw}. 

The classical complexity class \NP\ contains all decision problems where
for ``yes'' instances, an efficiently checkable proof can be found, as 
e.g.\ colorability of a graph. A problem is said to be \emph{hard} for a
class if any problem in this class can be reduced to solving this very
problem, and \emph{complete} if it is additionally inside the class.
While \NP-hardness of a problem strictly speaking does not prove that it
cannot be solved efficiently, it is the best we can hope for, given that
showing whether $\NP\ne\textbf{P}$, although generally believed to be
true, is one of the most important open questions in complexity theory.

To construct hard ground state problems for DMRG, we start from the
\textsc{local hamiltonian} problem and the corresponding complexity class
\QMA\ (Quantum Merlin Arthur).  \QMA\ contains those decision problems
where for ``yes'' instances, there is a quantum proof which can be checked
efficiently by a quantum computer, and is thus the natural quantum
generalization of \NP.
 More precisely, in the definition of
\QMA\ there are thresholds $p>q$: for ``yes'' instances, there is a proof
which will be accepted with probability at least $p$, whereas for ``no''
instances, no attempt to provide a fake proof will succeed with
probability more than $q$.  It is sufficient to require $p-q>1/\poly(N)$
(with $N$ the problem size), since then, the probabilities can be
amplified up to exponentially close to $1$ and $0$,
respectively~\cite{aharonov:quantum-np}.  As shown by
Kitaev~\cite{kitaev:lh,aharonov:quantum-np}, the problem \textsc{local
hamiltonian} is complete for \QMA:  Given a local Hamiltonian on $N$ spins
(where in this case \emph{local} means it is a sum of few-particle terms),
decide whether the ground state energy is below $a$ or above $b$, with
$b-a>1/\poly(N)$. It is easy to see that this problem is in \QMA---the
proof is the ground state (or several copies thereof), and the verifier
estimates the ground state energy by measuring the local terms.

Let us now review Kitaev's construction for proving \QMA-hardness of
\textsc{local hamiltonian}.  The task is,  given a polynomial-size 
quantum circuit (the verifier), to construct a local Hamiltonian
for which the ground state energy is at least $1/\poly(N)$ lower
if there exists a satisfying
input to the circuit, i.e., a valid proof.  To this end, write the verifying
circuit using $T=\poly(N)$ one- and two-qubit gates $U_t$, and for each
valid input $\ket{\phi_0}$ to the circuit construct a state which encodes the
history of the verifier checking this very input,
\begin{equation}
\label{qma-hist}
\ket{\psi}=\sum_{t=0}^T U_t\cdots U_1\ket{\phi_0\otimes0\cdots0}_d
    \ket{t}_t\ ,
\end{equation}
where $d$ denotes the data register [initially, the first part holds the
input $\ket{\phi_0}$ and the second $\poly(N)$ ancillas which are
initialized to $\ket{0\dots0}$; it thus consists of $M=\poly(N)$
qubits] and $t$ the time register.  Now, construct a Hamiltonian which
penalizes wrong
proof histories, 
\begin{equation}
\label{qma-ham}
H=H_\mathrm{init}+H_\mathrm{evol}+H_\mathrm{final}\ ,
\end{equation}
where $H_\mathrm{init}=T\sum_{a}\ket{1}_a\bra{1}\otimes\ket{0}_t\bra0$
penalizes any ancilla $a$ which is not properly
initialized,
\[
H_\mathrm{evol}=\sum_{t}
-U_{t}\ket{t}\bra{t-1}-U_{t}^\dagger\ket{t-1}\bra{t}+\ket{t-1}\bra{t-1}
+\ket{t}\bra{t}
\]
ensures that the transistions between $t-1$ and $t$ in $\ket{\psi}$
are correct,
and $H_\mathrm{final}=\ket{0}_1\bra{0}\otimes\ket T_t\bra T$ penalizes the
state $\ket{0}$ on the very first qubit---it will be set to $\ket 1$ if
the circuit accepts the proof.  It has been shown that if there exists a
proof which will be accepted by the verifier with high probablity, the
ground state energy of (\ref{qma-ham}) is by $1/\poly(N)$ lower than if
there is no such proof. The intuition is that in the former case, the
state (\ref{qma-hist}) almost does the job, while in the latter case, at
least one of the terms in the Hamiltonian (or a superposition thereof) has
to be violated.

In general, the spectral properties of (\ref{qma-ham}) will be complicated
since from to the definition of \QMA, there can be many potential
witnesses with different acceptance probabilities.  In the
following, we will restrict to problems where inputs to the
verifier are either accepted of rejected deterministically if choosen from
the proper basis, and this will simplify the spectrum remarkably.
Actually, we go even further and consider classical deterministic
verifiers, corresponding to problems in the complexity class
\NP~\footnote{ 
Note that the idea to use \emph{classical} verifiers, corresponding to the
class \MA, has already been used in \cite{stoquastic-ham} to show that
finding ground states of so-called \emph{stoquastic} Hamiltonians is
\MA-hard.  However, while this makes the Hamiltonian stoquastic, it does
not lead to a simple spectrum or weakly entangled proofs, since the
outcome is non-deterministic and $\ket{+}$ ancillas are being used to
obtain randomness.   Note that our Hamiltonians are also stoquastic if one
allows for a direct implementation of the Toffoli gate.
}, which will finally give rise to the simple entanglement structure
of the ground state we are after.

To determine the spectral properties of (\ref{qma-ham}) for a classical
deterministic circuit, let us fix a classical initial state of the data
register
$\ket{\bm{a}}_{d}=\ket{a_1\dots a_{M}}_{d}$ 
and analyze the system on the
$T+1$-dimensional space 
$\mathcal H_{\bm a}$
spanned by $\ket{\chi_t}=U_t\cdots U_1\ket{\bm
a}_{d}\ket{t}_t$, which is closed under the action of $H_\mathrm{init}$,
$H_\mathrm{final}$ and $H_\mathrm{evol}$. 
In particular, 
\begin{equation}
\label{subspace-ham}
H_{\bm a}
    =H\big|_{\mathcal H_{\bm a}}
    =\left(\begin{array}{ccccc}
	TA + 1 & -1 \\
	-1     & 2  & -1 \\
	       & \ddots & \ddots &\ddots \\
	       &        & -1 & 2 & -1 \\
	       &        &    & -1 & B+1 \\ 
    \end{array}\right)
\end{equation}
with $A=0,1,2\dots$ the number of wrongly initialized ancillas in
$\ket{\bm a}_{d}$ and $B=0,1$ depending on
whether the circuit accepts or rejects the input $\ket{\bm a}_{d}$. 
In the case of a ``yes'' instance of the \NP\ problem, there exists an
$\ket{\bm a}_{d}$ for which $A=B=0$, whereas for ``no''
instances, the lowest-reaching subspace has $A=0$, $B=1$.
In both cases, the eigenfunctions are
\begin{equation}
\label{log-eigenstates}
\ket{\psi_{\bm a,n}}= C \sum_{t=0}^T
    \cos[\omega_n(t+\tfrac12)]\ket{\chi_t}\;,\ n=0,\dots,T\;,
\end{equation}
where $\omega_n\equiv\omega_n^0=n\pi/(T+1)$ for $A=0$, $B=0$ and 
$\omega_n\equiv\omega_n^1=(n+\tfrac12)\pi/(T+\tfrac32)$ for $A=0$,
$B=1$, respectively; in both cases, the eigenvalues are given by 
$\lambda_n=2(1-\cos \omega_n)$, and $C^2=\Theta(1/T)$~\footnote{
$O$, $\Omega$, and $\Theta$ denote lower, upper, and exact 
bounds on the asymptotic scaling, respectively.}.

Different from \QMA-completeness proofs, we are not interested in
the difference in ground state energy between ``yes'' and ``no'' instances
but rather in the spectral gap for each of the cases independently.
Analyzing the spectrum will be simplified a lot by the fact that the
subspaces $\mathcal H_{\bm a}$ are closed under the action of any term in
the Hamiltonian, which is due to the restriction to classical
deterministic circuits.

We start by analyzing $H$ for a circuit corresponding to a ``yes''
instance of the \NP\ problem.  Then, there is at least one initial state
$\ket{\bm a_0}_{d}$
such that $H_{\bm a_0}$ has $A=B=0$ in
(\ref{subspace-ham}), and since $H\ge0$, this subspace contains a ground
state. There are two different types of excited states: The ones within
$\mathcal H_{\bm a_0}$, which have a gap
$2(\cos\,\omega_0^0-\cos\,\omega_1^0)=\Omega(1/T^2)$, and those 
within another $\mathcal H_{\bm a}$ for which $A$ and/or $B$ are strictly
positive.  The energy in any of these subspaces is bounded by the
ground state energy for $A=0$, $B=1$, and is thus $\Omega(1/T^2)$ as well,
proving an $\Omega(1/T^2)$ spectral gap of the overall Hamiltonian. Note
that the degeneracy of the ground state manifold equals the number of
different accepted inputs. 

On the other hand, for ``no'' instances there is no subspace with $A=B=0$.
It is easily seen that the ground state subspace has $A=0$, $B=1$, with
ground state energy $2(1-\cos\,\omega^1_0)$ and an
$\Omega(1/T^2)$ gap within the subspace. In order to bound the gap to
subspaces with $A\ge1$ (for which $B$ can be $0$), we use the following
lemma, shown in~\cite{kitaev:lh}: Given 
finite-dimensional operators $P\ge0$, $Q\ge0$ with null eigenspaces,
\begin{equation}
\label{gaplemma}
P+Q\ge \min\{\Delta(P),\Delta(Q)\}(1-\cos\,\theta)\ ,
\end{equation}
where  $\Delta(O)>0$ is the smallest nonzero eigenvalue of $O$,
and $\theta$ the angle between the null spaces of $P$ and $Q$.
It follows that the lowest eigenvalue in an $A=1$, $B=0$ subspace is
at least $T(1-\cos\,\omega_1^0)(1-\sqrt{T/(T+1)})$, and thus any subspace
with $A\ge1$ has an energy $\Omega(1/T^2)$ above the ground state.

An important point to observe is the particularly simple entanglement
structure of the eigenstates (\ref{log-eigenstates}) of $H$,
 which in 1D will
allow to represent them as MPS.  To see this, take any classical
reversible
verifying circuit and decompose it into a sequence of local gates.  Let us
first allow for three-qubit gates, so we can use the Toffoli gate which is
universal for classical reversible computation. Since it is a classical
gate, each of the states $\ket{\bm a(t)}=U_t\cdots U_1\ket{\bm a}_{d}$ is
classical itself, and thus each of the eigenstates
(\ref{log-eigenstates}) is a superposition of only $T+1$ classical
terms $\ket{\bm a(t)}\ket t\equiv\ket{\chi_t}$. 
As we want to restrict to two-qubit gates, each Toffoli gate has to be
implemented using a short sequence of entangling two-qubits gates. 
This temporarily creates entanglement
between the three neighboring qubits on which the Toffoli is 
applied, thus adding some entanglement to the eigenstates.
However, this entanglement is both spatially and ``temporally''
restricted, since it only involves three qubits and it only persists over
a few timesteps.

Let us now turn towards one-dimensional systems, where we will employ the
one-dimensional \QMA-complete Hamiltonian construction of Aharonov
\emph{et al.}~\cite{aharonov-irani:1d-qma}.  Since the restriction to
one-dimensional local Hamiltonians makes it impossible to access a global
time register, the time is encoded in the position of the data
register.  To this end, the data register is realized sequentially $T+1$
times in the 1D system [which thus consists of $L=M(T+1)=\poly(N)$ sites],
and with each timestep, the active register moves to the right.  To mark
which is the active register and to implement a
Hamiltonian (\ref{qma-ham}), a control register is appended to each qubit.
It is used both to store the status of the register (i.e.,
used/active/unused) and to implement an involved scheme in which a head is
moving back and forth, thereby first implementing the desired operation
$U_t$ on the active register and then, qubit by qubit, copying it to the
next timeslice. Thereby, each original $t$-timestep is encoded in
$K=O(M^2)$ elementary movements of the head (``$\tau$-timesteps''),
replacing the original $T$ steps by $\mathcal T=KT=\poly(N)$ steps of the
encoded system. The resulting local dimension per site is $12$ (achieved by
removing the data qubit degree of freedom e.g.\ for non-active blocks), and the
resulting Hamiltonian acts on nearest neighbors only.

As before, the Hamiltonian is a sum of transition rules $\tilde
H_\mathrm{evol}$ (now encoding the elementary movements of the head) and
of penalties $\tilde H_\mathrm{init}$ and $\tilde H_\mathrm{final}$ for
undesired initial and final configurations, acting on the blocks
corresponding to $t=0$ and $t=T$, which are applied when the head moves
over the qubit.  Additionally,  one now has to make sure the system stays
in the subspace of allowed configurations of the status register,
excluding e.g.\ the occurence of more than one head.
 This is achieved by adding a sum of
local penalty terms $\tilde H_\mathrm{penalty}$ acting on the status
register, which either penalize forbidden configurations directly or
indirectly as they evolve to penalized ones under $\tilde
H_\mathrm{evol}$.  For details on the implementation, see~%
\cite{aharonov-irani:1d-qma}.

To analyze the spectral properties of the 1D Hamiltonian, we apply the
``clairvoyance lemma'' of Ref.~\cite{aharonov-irani:1d-qma} which tells us
that we can restrict our attention to the subspace of valid
configurations of the status register. Therefore, split the the total
Hilbert space into subspaces $\mathcal K_S$ spanned by minimal sets of
classical status register configurations $\mathcal S$ closed
under $\tilde H_\mathrm{evol}$; data degrees of freedom are 
left unrestricted. By
definition, these subspaces are also closed under $\tilde
H_\mathrm{penalty}$.  There is one subspace $\mathcal K_0$ which contains
only valid configurations, whereas all other $\mathcal K_S$ contain 
only illegal configurations.  The clairvoyance lemma shows that although
some of these configurations might not be directly detected by
$\tilde H_\mathrm{penalty}$, the minimal energy of
$\tilde H_\mathrm{evol}+\tilde H_\mathrm{penalty}$, restricted to any of these
subspaces, is $\Omega(1/\mathcal T^3)$, and since $\tilde H_\mathrm{init}\ge0$ and
$\tilde H_\mathrm{final}\ge0$ act on the data register and thus within the
subspace, they do not affect this lower bound. By multiplying $\tilde
H_\mathrm{penalty}$ by $\mathcal T^2$, we can boost this to
$\Omega(1/\mathcal T)$ which will be sufficiently above the
low-lying eigenstates in $\mathcal K_0$.

On the subspace $\mathcal K_0$, $\tilde H_\mathrm{penalty}$ vanishes and
we can proceed as before: We choose an initial classical configuration
$\bm a$ of the data register and consider the system on the resulting
subspace $\mathrm{span}\,\{\ket{\chi_0},\dots,\ket{\chi_{\mathcal T}}\}$.
There, it is described by a Hamiltonian very similar to
(\ref{subspace-ham}), except for minor differences in the implementation
of $\tilde H_\mathrm{init}$:  While $\tilde H_\mathrm{final}$ can well be
applied in the very last
$\tau$-timestep on the rightmost data qubit and thus give the same
penalty term $B=0,1$ in (\ref{subspace-ham}), the penalties enforcing
properly initialized ancillas can appear in the first $M$
$\tau$-timesteps, i.e., a penalty $\mathcal T$ can show
up in any of the first $M$ diagonal entries of (\ref{subspace-ham}).
Since this only increases $\theta$ in (\ref{gaplemma}) and thus the gap,
we obtain the same spectral properties as before (but
we alo have to use the lemma for the ``yes'' instances). 
Note that
in particular, all energies are well below the energy of any subspace
$\mathcal K_S$ with illegal configurations.

Let us now investigate the entanglement structure of the low-lying
eigenstates which are of the form (\ref{log-eigenstates}), but in the
one-dimensional encoding.  As before, there are two sources of
entanglement: On the one side, we have a superposition of all
$\tau$-timesteps, i.e., of $\mathcal T=\poly(N)$ states.  Each of them
is almost classical, with the only source of entanglement being
\emph{one} Toffoli gate which is performed.  As this involves three qubits,
each of these states is a superposition of at most $8$ classical states,
and thus all low energy eigenstates of the one-dimensional Hamiltonian can
be written as a superposition of $\poly(N)$ classical states.  It follows
immediately that the ground state can be represented by an MPS with bond
dimension $D=\poly(N)$.  Note that the bond dimension actually needed is
considerably smaller than the number of classical states, since e.g.\ the
superposition of $O(M^2)$ states which arises from encoding the
$t\leftrightarrow t+1$ transition in $O(M^2)$ $\tau$-timesteps ranges over
two consecutive $t$-timeslices and thus $2M$ sites only.

We have shown that if the one-dimensional \QMA\ construction of
\cite{aharonov-irani:1d-qma} is applied to problems from the class \NP,
the resulting Hamiltonian has a polynomial gap, and the low-lying
eigenstates are MPS.  Let us now see what this implies for the difficulty
of finding ground states of one-dimensional systems. Firstly, let us
encode the verifying circuit for an \NP-\emph{complete} problem in the
Hamiltonian. Thereby, we obtain a polynomially gapped Hamiltonian for
which the ground state manifold is spanned by MPS, and for which finding a
ground state---or even an approximation within an accuracy sufficiently
smaller than the gap---is an \NP-hard problem. Note that obtaining the MPS
representation of the ground state is indeed stronger than just deciding
the \NP\ problem itself, since from it one can efficiently extract the
satisfying assignment.

Let us now construct Hamiltonians with a unique ground state.  In order to
have a unique ground state for all instances, we have to restrict to
problems which do not only have unique proofs for ``yes'' instances,  but
also unique disproofs for ``no'' instances, or more formally problems in
$\NP\cap\coNP$ with unique proofs (\coNP\ is the class of problems where
``no'' instances can be disproven).  Although this is likely to be a
smaller class than $\NP$, it still contains interesting hard
problems~\cite{papadimitriou:book}.  In particular, finding the prime
factor decomposition of a given number corresponds to a problem in this
class: The prime factor decomposition always exists, it is unique, and
since primality testing is in \P, it can be efficiently checked whether a
given decomposition is indeed the prime factor decomposition.  The
corresponding Hamiltonian has an MPS as its unique ground state, a
polynomial gap above it, it is frustration free since there is always an
accepting input, and approximating the ground state by an MPS is at least
as hard as factoring;  in particular, the prime factor decomposition
itself can be read right off the MPS representation of the ground state.

Beyond the direct implications on the hardness of finding MPS ground
states, our results also provide strong evidence against the existence of
a certifyable version of DMRG.  The idea behind \emph{certifyability} is
that even if an algorithnm does not always converge to the true ground
state, in the cases where it succeeds this can be certified.  To see why
no variational method over MPS can be certifyable, take again an
\NP-complete problem encoded in a one-dimensional Hamiltonian. On the one
hand, for any ``yes'' instance success can be readily checked as the
Hamiltonian is frustration free.  However, if it was possible to certify
the ground state for ``no''-instances, this would be a way to provide an
efficiently checkable certificate that one is facing a ``no''-instance
for any problem in \NP, and would therefore prove $\NP=\coNP$, which is
considered unlikely. 

Finally, since the Hamiltonian for a ``yes''-instance is always
frustration free, one cannot even hope for a DMRG algorithm which only
works for frustration free Hamiltonians.  Otherwise, run the algorithm on
any instance of an \NP\ problem and check the energy of the state
returned: for a ``yes'' instance, the Hamiltonian is frustration free and
thus the energy is $0$, whereas ``no'' instances can be easily detected
due to their larger ground state energy.  Thus, the existence of such an
algorithm would allow to solve \NP-complete problems in polynomial time.

Note that our construction to obtain \NP-hard ground state problems with
simple spectral properties will work for any \QMA\ scheme, as 
all of them are based on Kitaev's original construction.  For
two-dimensional Hamitonians this is less interesting since
there exist hard classical Hamiltonians; however it might be 
interesting to apply it to the translational invariant constructions
of~\cite{tinv-lh}.

\emph{Acknowledgements.}---%
We thank S.\ Bravyi, A.~Kay, and T.~Osborne for helpful discussions.  N.~S.\ thanks the
Erwin Schr\"odinger Institute in Vienna, where parts of this work
were carried out, for their hospitality.  This work has been supported by
the EU (SCALA) and by the cluster of excellence project MAP.

\end{document}